\documentclass{ws-procs961x669}
      
\begin{document}
\title{Breaking Buchdahl: ultracompact stars in semiclassical gravity}

\author{Julio Arrechea$^*$, Carlos Barceló$^\dagger$}

\address{Institute of Astrophysics of Andalusia (IAA-CSIC), \\
Glorieta de la Astronomía, 18005 Granada, Spain\\
$^{*}$E-mail: arrechea@iaa.es,~ $^\dagger$E-mail: carlos@iaa.es}

\author{Raúl Carballo-Rubio}

\address{Florida Space Institute, University of Central Florida,\\
12354 Research Parkway, Partnership 1, 32826 Orlando, FL, USA\\
E-mail: Raul.CarballoRubio@ucf.edu}

\author{Luis J. Garay}

\address{Universidad Complutense de Madrid, 28040 Madrid, Spain\\
and\\
Instituto de Estructura de la Materia (IEM-CSIC),\\
 Serrano 121, 28006 Madrid, Spain\\
E-mail: luisj.garay@ucm.es}

\begin{abstract}
The semiclassical approximation takes into account the gravitational contribution of zero-point energies. We model this contribution via the renormalized stress-energy tensor (RSET) of a massless scalar field, which we compute in a cutoff-regularized version of the Polyakov approximation. When the field is in the Boulware vacuum state (the natural vacuum for stellar geometries), the RSET works in favor of violating the Buchdahl compactness limit. We review the family of classical constant-density stellar solutions, paying particular attention to the notion of criticality---the presence of offsets in the mass function---and use it as a warm up for the analysis of the semiclassical set of solutions. For stars that surpass Buchdahl limit by far, the critical solution has an irregular pressure. This divergence in pressure moves inward by introducing a negative offset in the mass. In the semiclassical theory we find something rather different, namely 
that the critical configuration already displays a pressure that diverges exactly at the center of the structure. This drastic difference between the classical and semiclassical space of solutions suggests that semiclassical gravity could potentially allow for the existence of ultracompact stellar objects. 
\end{abstract}


\bodymatter

\section{Introduction}\label{Sec: Introduction}
In recent times, new windows have opened towards probing the nature of astrophysical black holes through both gravitational wave \cite{Abbottetal2016} and electromagnetic \cite{Akiyamaetal2019} observations. As observations aspire to unveiling the nature of these dark and compact objects, the issue of guessing what lurks behind their event horizons, assuming these long-lived surfaces exist in the first place, becomes more pressing than ever~\footnote{Expanded version of the contributed talk by J.A. for the Sixteenth Marcel Grossmann Meeting (2021).}. Among all the different theories that, in principle, manage to change our notion of theoretical (Schwarzschild) black holes, semiclassical gravity stands out, mainly because it exhibits a conceptually well-stablished framework that has led to the discovery of effects such as cosmological particle creation \cite{Parker1968} and Hawking evaporation \cite{Hawking1976}.

The main idea behind semiclassical gravity is that an effective geometrical description of the spacetime is retained, while the sources of this classical spacetime are of quantum origin. Therefore, this approximate scheme could be regarded as a conservative approach towards a theory of quantum gravity, in the sense that it captures the first corrections to classical general relativity originated by the zero-point energies of quantum matter fields. The presence of spacetime curvature modifies these zero-point energies in a way that cannot be renormalized away, the subsequent contribution from these zero-point energies to spacetime curvature being captured by the semiclassical field equations
%----------------------------------------------------------
\begin{equation}\label{Eq:SemiEinstein}
G_{\mu\nu}=8\pi\left(T_{\mu\nu}+\hbar\langle\hat{T}_{\mu\nu}\rangle\right),
\end{equation}
%----------------------------------------------------------
where $\langle\hat{T}_{\mu\nu}\rangle$ is the vacuum expectation value of the renormalized stress-energy tensor (RSET) of quantized fields. Obtaining analytical expressions for the RSET is an arduous task \cite{Wald1978}, so it is typical to resort to highly symmetrical or dimensionally reduced models where the RSET can be computed explicitly \cite{BrownOttewill1985, Frolov1987}. 

%----------------------------------------------------------
\section{The Polyakov RSET: Boulware vs Buchdahl}\label{Sec:Polyakov}
%----------------------------------------------------------

For simplicity, we restrict our analysis to static and spherically symmetric spacetimes with line element
\begin{equation}\label{Eq:Metric}
ds^{2}=-e^{2\phi(r)}dt^{2}+\left[1-C(r)\right]^{-1}dr^{2}+r^{2}d\Omega^{2},
\end{equation}
where $d\Omega^{2}$ is the line element of $2$-spheres and $\phi,~C$ are the redshift and compactness functions, respectively. The former measures the amount of redshift suffered by outgoing null rays and the latter is a measure of the amount of mass contained within concentrical spheres, namely $C(r)=2m(r)/r$, with $m(r)$ the Misner-Sharp mass \cite{MisnerSharp1964}.

An appealing method for obtaining a particularly simple RSET is via the $s$-wave Polyakov approximation \cite{Polyakov1981}. This approximation models the RSET as that of a massless scalar field propagating in a $1+1$ dimensional spacetime [the $t,r$ sector of \eqref{Eq:Metric}]. The resulting RSET is a function of the components of the metric \cite{DaviesFulling1977} and has the form
%----------------------------------------------------------
\begin{align}
    \langle\hat{T}_{rr}\rangle^{(2)}=
    &
    -\frac{l_{\rm P}^{2}\psi^{2}}{2}+\langle\rm{SDT}\rangle,\quad \langle\hat{T}_{rt}\rangle^{(2)}=\langle\hat{T}_{tr}\rangle^{(2)}=0,\nonumber\\
    \langle\hat{T}_{tt}\rangle^{(2)}=
    &
    \frac{l_{\rm P}^{2}e^{2\phi}}{2}\left[2\psi'\left(1-C\right)+\psi^{2}\left(1-C\right)-\psi C'\right]+\langle\rm{SDT}\rangle,
\end{align}
%----------------------------------------------------------
where $\langle\rm{SDT}\rangle$ stands for state dependent terms, $l_{\rm P}=1/\sqrt{12\pi}$, $\psi=\phi'$ and the $'$ denotes derivatives with respect to the $r$ coordinate. These components are then identified with those of a $3+1$ dimensional tensor
%----------------------------------------------------------
\begin{equation}\label{Eq:PolyakovApprox}
\langle\hat{T}_{\mu\nu}\rangle^{\left(\rm P\right)}=F(r)\delta^{a}_{\mu}\delta^{b}_{\nu}\langle\hat{T}_{ab}\rangle^{\left(2\right)},
\end{equation}
%----------------------------------------------------------
where Latin indices take $0,1$ values and Greek indices take $0,3$ values. 

The Polyakov RSET provides an appropriate balance between simplicity and accuracy (see \cite{Arrecheaetal2021b} for an extended discussion). There is an ambiguity in the $3+1$-dimensional Polyakov RSET given by the radial function $F(r)$. Typically, this function is fixed to be $F(r)=1/r^2$ to yield a covariantly conserved RSET with vanishing angular pressures. In turn, the resulting RSET is singular at $r=0$, even computed over entirely regular stellar spacetimes. This poses a problem when backreaction is taken into account and it is the main motivation behind exploring other choices of regulator function. Following a conservative logic, we supplied the regulator with a cut-off of the form $F_{\rm CP}(r)=1/(r^{2}+\alpha l_{\rm P}^{2})$, with $\alpha>1$. This distortion of the regulator function induces a non-conservation of the RSET, which can be compensated by the introduction of angular pressures \cite{Arrecheaetal2020}. 

In previous works \cite{Arrecheaetal2020, Arrecheaetal2021} we explored the sets of semiclassical vacuum and electrovacuum solutions using the cut-off regulator Polyakov RSET. In what follows we will consider the Boulware vacuum state, obtained by taking $\langle\rm{SDT}\rangle=0$, as it is the sole state compatible with staticity (absence of stationary fluxes) and asymptotic flatness. The Boulware state is singular at the Schwarzschild event horizon, hence backreaction will affect it in a non-perturbative manner. The semiclassical counterpart of the Schwarzschild black hole has its horizon replaced by the throat of an asymmetric wormhole. At finite affine distance inside this throat lives an undressed singularity. This result motivates the introduction of a classical matter fluid as the only semiclassically-consistent possibility of obtaining geometries devoid from singularities. We will explore whether these geometries are potentially able to mimic astrophysical black holes.

This potentiality comes from realizing that semiclassical physics provides several ways of surpassing the Buchdahl compactness bound \cite{Buchdahl1959} that applies to relativistic stars in hydrostatic equilibrium. The conditions for deriving this limit involve that (i) the star has a Schwarzschild exterior, (ii) pressures in the angular directions that do not surpass the pressure in the radial direction, and (iii) a density profile that is non-increasing outwards. Semiclassical corrections modify the exterior (vacuum) geometry so that it is no longer Schwarzschild but an asymmetric wormhole. The RSETs are anisotropic by construction, although this anisotropy is underestimated in regularized Polyakov-like approximations. Lastly, the violations of energy conditions provided by the RSET generate regions inside the star where the energy density decreases inwards. 

In this contribution, we report on several results concerning the self-consistent solutions to \eqref{Eq:SemiEinstein}. We will point out the remarkable differences that exist between constant-density stellar solutions in the classical and semiclassical theories. We will elaborate on the notion of criticality---a criteria for identifying stellar geometries that exhibit deficit or excess of mass that come from singular sources---and use it to obtain a family of \textit{quasi-regular} semiclassical geometries. These quasi-regular geometries are characterized by a central (singular) core such that at its boundary the mass function is not trans-Planckian and pressures are finite. 

%----------------------------------------------------------
\section{Criticality in classical stars}\label{Sec:Criticality}
%----------------------------------------------------------
First we give a brief summary of classical stellar equilibrium applied to the constant-density perfect fluid. In this simplified setting, the concept of criticality can be introduced in a clear fashion, paving the path toward the more involved analysis involving semiclassical corrections.
The stress-energy tensor (SET) of the isotropic perfect fluid is
%----------------------------------------------------------
\begin{equation}\label{Eq:ClassicalSET}
    T_{\mu\nu}=\left(\rho+p\right)u_{\mu}u_{\nu}+pg_{\mu\nu},
\end{equation}
%----------------------------------------------------------
with $p$ and $\rho$ denoting the pressure and energy density measured by an observer comoving with the fluid with $4$-velocity $u^{\mu}$. In the following we will consider that the fluid obeys an equation of state of the form
%----------------------------------------------------------
\begin{equation}\label{Eq:State}
\rho=\text{const}.
\end{equation}
There are several reasons behind considering a constant-density fluid. First, this equation of state leaves the pressure of the fluid free to arrange itself in whichever form necessary to attain equilibrium. Second, the classical field equations admit (in some cases) analytical solutions. This characteristic is lost once we introduce semiclassical corrections. 
Finally, the constant-density assumption saturates one of the conditions necessary for the Buchdahl limit to hold~\cite{UrbanoVeermae2018}.

 Covariant conservation of the classical SET yields
%----------------------------------------------------------
\begin{equation}\label{Eq:Cont}
    p'=-\left(\rho+p\right)\psi.
\end{equation}
%----------------------------------------------------------
The $rr$ and $tt$ components of the classical field equations \eqref{Eq:SemiEinstein} take the form
%----------------------------------------------------------
\begin{equation}\label{Eq:Componentsrr}
     C=
     \frac{-8\pi r^{2}p+2r\psi}{1+2r\psi},\quad
    C'=
    8\pi r \rho-\frac{C}{r}.
    \end{equation}
%----------------------------------------------------------
Expressions (\ref{Eq:State},~\ref{Eq:Cont},~\ref{Eq:Componentsrr}) form a closed system. The differential equation for $C$ can be directly integrated, yielding
%----------------------------------------------------------
\begin{equation}
C=\frac{8}{3}\pi r^{2}\rho+\frac{2m_{0}}{r}.
\end{equation}
%----------------------------------------------------------
Here, $m_{0}$ is an arbitrary integration constant that accounts for a constant shift in the Misner-Sharp mass $m(r)$. The following relation
%----------------------------------------------------------
\begin{equation}\label{Eq:CriticalityClas}
M_{\text{ADM}}=\int_{0}^{R}4\pi r^{2}\rho~\text{d}r+m_{0}.
\end{equation}
%----------------------------------------------------------
calls for interpreting $m_{0}$ as a term that compensates the disagreement between the mass generated by the whole fluid sphere and the ADM mass $M_{\text{ADM}}$. Therefore, regularity in the compactness of a classical star enforces $\rho=\rho_{\text{crit}}=3C(R)/8\pi R^{2}$, where $\rho_{\text{crit}}$ stands for the critical density. We use this value to distinguish three families of solutions
\begin{itemize}
\item Sub-critical regime $\left(\rho<\rho_{\text{crit}}\right)$: These solutions are characterized by a positive $m_{0}$ and can be interpreted as a fluid sphere with a mass excess whose gravitational pull cannot be withstood by finite pressures. Sub-critical solutions have infinite pressure singularities at some $r>0$. For a fixed $C(R)$, the radius where this singularity appears nears the surface as $\rho$ decreases.
\item Critical regime $\left(\rho=\rho_{\text{crit}}\right)$: These solutions have a vanishing mass at $r=0$ $\left(m_{0}=0\right)$ and can be regular or irregular depending on whether the Buchdahl compactness bound is being surpassed. Critical solutions that surpass Buchdahl limit $(C(R)=8/9)$ display a pressure divergence at
%----------------------------------------------------------
	\begin{equation}
	R_{\text{div}}=3R\sqrt{1-\frac{8}{9C(R)}}.
	\end{equation}
%----------------------------------------------------------
\item Super-critical regime $\left(\rho>\rho_{\text{crit}}\right)$: Stellar geometries characterized by a negative $m_{0}$. The gravitational repulsion exherted by this negative mass tames the growing of the pressure of the fluid but at the cost of producing a curvature singularity
at $r=0$. As $m_0$ becomes more negative, the pressure at the core decreases, eventually becoming finite at $r = 0$ for some separatrix density.
\end{itemize}

\begin{figure}
    \centering
    \includegraphics[width=\columnwidth]{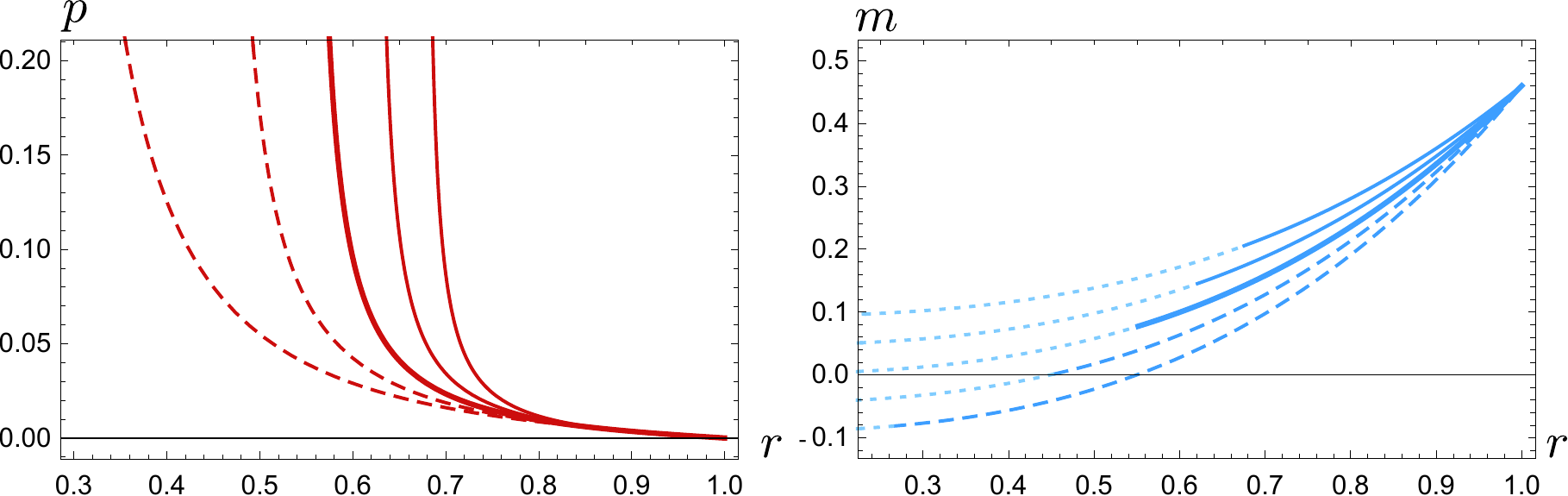}
    \caption{Pressure (left panel) and Misner-Sharp mass (right panel) profiles surrounding the critical solution of a star with~$C(R)=0.92$ and $R=1$. The thick, continuous line denotes the critical solution, while thin lines are sub-critical and dashed lines are super-critical stars. The only solution regular in the mass (critical) has a highly irregular pressure.
The mass curves have been analytically extended for pictorical purposes, but the associated pressure profiles are singular. 
From right to left (top to bottom) $\rho/\rho_{\text{crit}}=\left\{0.8,0.9,1.0,1.1,1.2\right\}$. }
    \label{Fig:Clas_Crit}
\end{figure}

\begin{figure}
    \centering
    \includegraphics[width=\columnwidth]{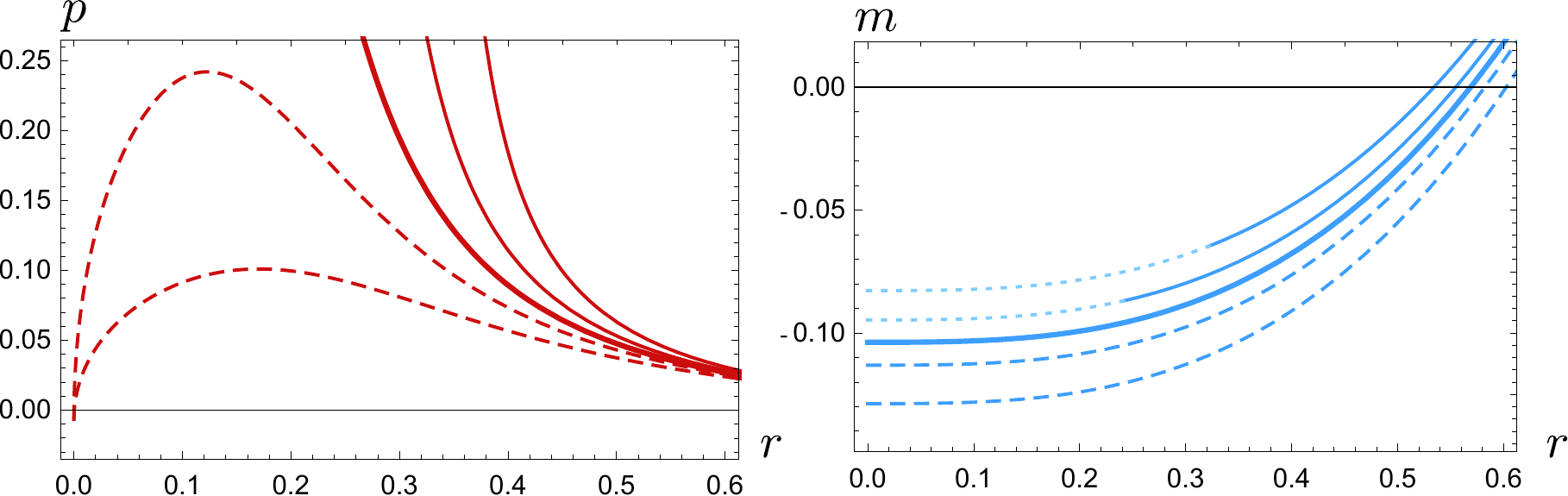}
    \caption{Pressure (left panel) and Misner-Sharp mass (right panel) profiles surrounding the pressure-separatrix solution of a star with $C(R)=0.92$ and $R=1$. The thick, continuous line denotes the separatrix solution, while thin lines are infinite-pressure solutions and dashed lines are finite-pressure ones. Note that making the pressure everywhere finite enforces a negative central mass. All integrations are super-critical because $m_{0}<0$. 
From right to left (top to bottom) $\rho/\rho_{\text{crit}}\simeq\left\{1.18,1.21,1.23,1.25,1.28\right\}$. }
    \label{Fig:Clas_Sep}
\end{figure}

Despite its apparent simplicity, this space of solutions reveals an important characteristic: for stars that surpass the Buchdahl limit, $C(R)=8/9$, the respective separatrices for the pressure and mass functions do not overlap in any region.  Fig. \ref{Fig:Clas_Crit} shows integrations surrounding the critical solution, the only one with a regular compactness function\footnote{Recall that $C=2m/r$, so only a Misner-Sharp mass that vanishes at the origin returns a non-singular compactness.}.  The associated pressure profiles all develop singularities significantly far from $r=0$. On the other hand, Fig. \ref{Fig:Clas_Sep} shows integrations surrounding the solution separatrix in pressure, that is, the solution which is super-critical enough as to have a pressure that diverges exactly at $r=0$. However, the price to pay for generating finite-pressure configurations is that the mass goes highly super-critical. At this point, enforcing the mass function to vanish at $r=0$ would require introducing an additional positive mass contribution inside a small core. This sort of regularization cannot come without additional pathologies as the density of such sphere would necessarily be trans-Planckian. In what follows we will show that semiclassical corrections naturally provide this regularization mechanism for the mass, although they fail in achieving a strictly regular behavior at $r=0$.
%----------------------------------------------------------
%----------------------------------------------------------
%----------------------------------------------------------
%----------------------------------------------------------

%----------------------------------------------------------
\section{Semiclassical corrections in stellar spacetimes}\label{Sec:Semiclassical}
%----------------------------------------------------------
The classical equations \eqref{Eq:Componentsrr} guarantee the behavior of the compactness is independent from that of the pressure, although the converse is not true. The addition of semiclassical corrections intertwines the behavior of both functions. As a consequence, when we analyze the semiclassical equations, the notion of criticality becomes more involved. This is because the differential equation for the compactness cannot be integrated analytically, so we lack an explicit expression that informs about the value of $\rho_{\text{crit}}$.  The most relevant departure of semiclassical stars with respect to their classical counterparts is that, in the semiclassical theory, the separatrix solutions in mass and pressure overlap in a narrow region of the parameter space. In the following we will catalogue the space of semiclassical constant-density solutions and elaborate on this aspect.

Introducing the cut-off regularized Polyakov RSET \eqref{Eq:PolyakovApprox} as a curvature source in the field equations \eqref{Eq:SemiEinstein}, they result in
%----------------------------------------------------------
\begin{align}\label{Eq:SComponentsrr}
     C=
     &
     \frac{-8\pi r^{2}p+2r\psi+l_{\rm P}^{2}r^{2}\psi^{2}/\left(r^{2}+\alpha l_{\rm P}^{2}\right)}{1+2r\psi+l_{\rm P}^{2}r^{2}\psi^{2}/\left(r^{2}+\alpha l_{\rm P}^{2}\right)},\\
    C'=
    &
    \frac{8\pi r^{2} \rho-C+l_{\rm P}^{2}r^{2}\left(\psi^{2}+2\psi'\right)(1-C)/\left(r^{2}+\alpha l_{\rm P}^{2}\right)}{r+l_{\rm P}^{2} r^{2} \psi/\left(r^{2}+\alpha l_{\rm P}^{2}\right)}.\label{Eq:SComponentstt}
    \end{align}
%----------------------------------------------------------
Semiclassical corrections to stellar spacetimes are typically negligible (they are proportional to $\hbar$) unless the geometries under consideration are able to explore and stimulate the energetic contributions of the Boulware state as to compensate their suppression of order $\hbar$. The scale that can potentially compensate this screening of semiclassical effects is the compactness of the star. In the following, we will be considering spheres of fluid whose surface is located very close and above the throat of the vacuum wormhole geometry [$C(R)\to1$].
Such a configuration is defined by the following boundary conditions
%----------------------------------------------------------
\begin{equation}\label{Eq:BoundConds}
p(R)=0,~~~ \phi(R)=\phi_{\text{S}},~~~
\psi(R)=\frac{R^{2}+\alpha l_{\rm P}^{2}}{Rl_{\rm P}^{2}}\left[\sqrt{1+\frac{R^{2}}{R^{2}+\alpha l_{\rm P}^{2}}\frac{C(R)}{1-C(R)}}-1\right],
\end{equation}
%----------------------------------------------------------
and a given value of $\rho$ and $C(R)$. Here, the parameter $\phi_{\text{S}}$ is obtained through a numerical integration of the vacuum equations from the asymptotic region inwards.

Plugging the boundary conditions \eqref{Eq:BoundConds} in the field equations (\ref{Eq:SComponentsrr},~\ref{Eq:SComponentstt}) we obtain expressions for $\psi'(R)$ and $C'(R)$. Then, the semiclassical energy density and radial pressure at $r=R$ are
%----------------------------------------------------------
\begin{align}\label{eq:RSETvac}
\rho_{\rm{sc}}=
&
-\langle\hat{T}^{t}_{t}\rangle^{(\rm{P})} =
-\frac{1}{8\pi R^{2}} + \rho+\mathcal{O}\left(\sqrt{1-C}\right),\nonumber\\
p^{r}_{\rm{sc}}=
&
\langle\hat{T}^{r}_{r}\rangle^{(\rm{P})}=-\frac{1}{8\pi R^{2}}+\mathcal{O}\left(\sqrt{1-C}\right),\nonumber\\
p^{\theta}_{\rm{sc}}=
&
\langle\hat{T}^{\theta}_{\theta}\rangle^{\left(\rm P\right)}=-\frac{\alpha l_{\rm P}^{2}}{8\pi R^{2}\left(R^{2}+\alpha l_{\rm P}^{2}\right)}+\mathcal{O}\left(\sqrt{1-C}\right).
\end{align}
%----------------------------------------------------------
The RSET components are comparable in magnitude to the classical SET at the surface. In fact, as long as $\rho\lesssim1/8\pi R^{2}$ the total RSET violates energy conditions at the surface. Additionally, for more realistic equations of state with a classical energy density that vanishes at the surface, the total RSET will violate all pointwise energy conditions there.

Vacuum polarization provides an extra contribution to the Misner-Sharp mass of the spacetime. This contribution is negative in the vacuum region and permeates it entirely. In the interior of the star it can be overall positive or negative depending on the value of $\rho$. The equivalent to formula \eqref{Eq:CriticalityClas} in the semiclassical theory is
%----------------------------------------------------------
\begin{equation}\label{Eq:CritSemi}
M_{\rm{ADM}}=\int_{0}^{\infty}4\pi r^{2}\left[\Theta\left(R-r\right)\rho+\rho_{\rm sc}\right]dr +m_{0},
\end{equation}
%----------------------------------------------------------
where $\Theta$ is the Heaviside step function. Since the total energy density is now intertwined to the pressure and redshift of the geometry through Eq. \eqref{Eq:SComponentstt}, the critical value of the classical density (recall, the value that corresponds to a vanishing $m_{0}$) cannot be determined analytically. In turn, a numerical exploration of the space of solutions is needed. We dedicate the rest of this note to highlight the chief results from our exploration~\footnote{We refer the reader to \cite{Arrecheaetal2021b} for details.}.

Concerning the Buchdahl limit, we have found that the classical Buchdahl bound of $C(R)=8/9$ gets perturbatively modified by semiclassical corrections. These correction makes the most compact configuration which is regular in $p$ and $C$ have $C(R)<8/9$. This result follows from $\rho_{\rm sc}$ in \eqref{Eq:CritSemi} amounting to an overall positive contribution throughout the interior of the structure. In fact, the greatest positive contribution comes from the region $r\sim \sqrt{\alpha} l_{\rm P}$, precisely where the physics of the Polyakov RSET is driven by the regulator function. Using a local, more refined approximation to the RSET \cite{Hiscock1988} the energetic contribution near the center of constant-density stars that are approaching Buchdahl limit turns out to be negative. The cutoff-Polyakov RSET could be incorrectly estimating the magnitude and sign of the energetic contribution from the vacuum near the center of compact fluid spheres.

\begin{figure}
    \centering
    \includegraphics[width=0.6\columnwidth]{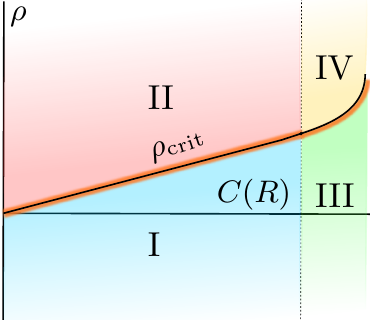}
    \caption{Pictorial representation of an $R=\text{const.}\gg l_{\rm P}$ slice of the semiclassical constant-density stellar space of solutions. The vertical and horizontal axes represent the energy density and surface compactness of stars. The black curve denotes critical configurations. Its intersection with the vertical, dotted line represents the Buchdahl limit, or the most compact structure regular in both $p$ and $C$. We distinguish four regions: regions $\rm{I}$ and $\rm{II}$ for sub- and super-critical stars below Buchdahl limit and regions $\rm{III}$ and $\rm{IV}$ for sub- and super-critical stars above Buchdahl limit. The narrow orange band that surrounds the $\rho_{\text{crit}}$ line denotes the quasi-regular regime which spans through regions $\rm{I}, \rm{II}$ and $\rm{III}$.}
    \label{Fig:Phase}
\end{figure}

In our exploration of the semiclassical space of solutions (see Fig. \ref{Fig:Phase} for a pictorical representation) we found the following regimes 
\begin{itemize}
\item Sub-critical regime ($\rho\ll\rho_{\text{crit}}$, regions $\rm{I}$ and $\rm{III}$ of the diagram in Fig.~\ref{Fig:Phase}): These solutions are asymmetric wormholes with characteristics akin to the vacuum solution, but filled with a classical fluid of diverging pressures. As $\rho$ is increased, the throat of the wormhole progressively shrinks. When the throat is very small compared to the overall size of the star, we identify a new regime not present in the classical theory.
\item Quasi-regular regime ($\rho\sim\rho_{\text{crit}}$, narrow orange band in Fig.~\ref{Fig:Phase}): This subfamily is comprised by super-critical solutions that lie below Buchdahl limit and sub-critical solutions of any compactness with an extremely tiny wormhole throat. In this regime, we can always identify a central core of radius $r_{\text{core}}\ll R$. At the boundary of this core the pressure remains finite and the Misner-Sharp mass can be made as small as desired. 
\item Critical regime ($\rho=\rho_{\text{crit}}$, black line in Fig.~\ref{Fig:Phase}): Sub-Buchdahl stars amount to a perturbative correction over the classical geometries and are regular. 
Super-Buchdahl stars are irregular but extend all the way to $r=0$, contrary to their classical counterparts. These solutions have infinite positive pressure and infinite negative mass at $r=0$.
\item Super-critical regime ($\rho>\rho_{\text{crit}}$, regions $\rm{II}$ and $\rm{IV}$ of the diagram in Fig.~\ref{Fig:Phase}): They correspond to perturbative corrections over the classical solutions, with the caveat that the Misner-Sharp mass is now infinite and negative at $r=0$, which strengthens the central singularity. Pressure is finite everywhere.
\end{itemize}
%----------------------------------------------------------
%----------------------------------------------------------

\begin{figure}
    \centering
    \includegraphics[width=\columnwidth]{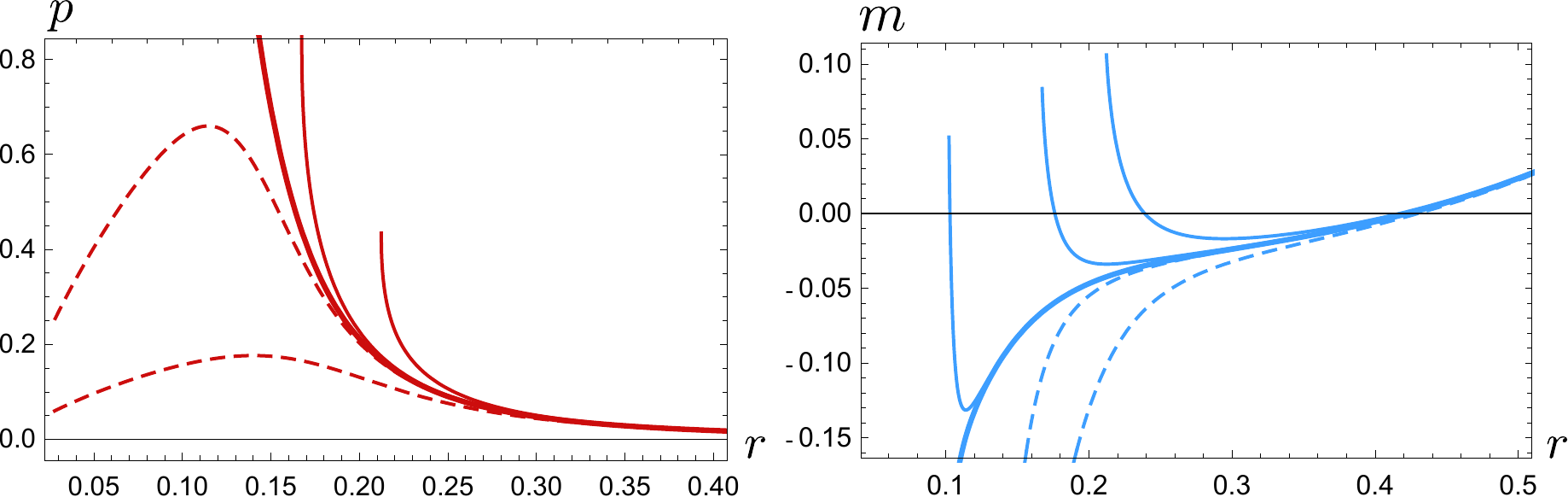}
    \caption{Pressure (left panel) and Misner-Sharp mass (right panel) profiles surrounding the critical solution of a semiclassical star with $C(R)=0.92, R=1$ and $\alpha=1.01$. The thick line denotes the critical solution (the separatrix between distinct compactness behaviors), while the thin lines are sub-critical and the dashed lines are super-critical. In the right panel the quasi-regular regime can be observed. Quasi-regular solutions are characterized by a mass function that vanishes in the innermost regions of the star. By a suitable choice of $\rho$, this vanishing can be achieved as close to $r=0$ as desired. From right to left (top to bottom) the densities utilized are \mbox{$\rho/\rho_{\text{crit}}=\left\{0.9990,0.9998,0.9999,1,1.0001,1.0010\right\}$}, where the critical density $\rho_{\text{crit}}\simeq0.1214$ is determined numerically within a precision of $10^{-6}$.}
    \label{Fig:Semi_Crit}
\end{figure}

Figure \ref{Fig:Semi_Crit} shows the pressure and mass profiles for various configurations surrounding the critical solution of a star that surpasses the Buchdahl limit (rightmost portion of diagram \ref{Fig:Phase}). There are several crucial differences between these diagrams and those from Figures \ref{Fig:Clas_Crit} and \ref{Fig:Clas_Sep}. Firstly, the critical solution (thick line), albeit singular, coincides with the separatrix solution for the pressure. In the classical case, the critical solution and the separatrix in the pressure were distant in the space of parameters. Here, because of the way quantum corrections operate, we find a coincidence between these two solutions. Secondly, quasi-regular configurations (thin lines) for which the Misner-Sharp mass vanishes near the center of spherical symmetry are very different from their classical counterparts. Enforcing the mass function to vanish at $r=0$ no longer requires the introduction of arbitrarily high masses inside small cores: Semiclassical corrections already provides a significant part of this contribution.
%----------------------------------------------------------
%----------------------------------------------------------

%----------------------------------------------------------
\section{Conclusions}\label{Sec:Conclusions}
%----------------------------------------------------------
In this work we have reviewed the classical stellar solutions of constant density under our particular scope, with the aim of understanding properly the solutions that incorporate vacuum polarization contributions. We have not been able to find strictly regular configurations that surpass a Buchdahl limit (which receives perturbative corrections). However, the study of singular solutions hints towards semiclassical gravity being able to regularize at the same time the pressure and mass of ultracompact configurations. This idea is what we have attempted to capture in the notion of quasi-regular configurations. 

We believe that the crudity of the approximations used here for the semiclassical source could be preventing the quasi-regular solutions from appearing as regular in the first place. Ideally, these analyses should be revisited when more accurate approximations to the RSET are available. As an intermediate step in this direction, it is possible to ask whether a deformation of the regulator function from its cutoff-regulator form $F_{\rm CP}$ just inside the central core of the structure is sufficient to generate entirely regular and ultracompact configurations. The results of this ongoing investigation will appear in a forthcoming publication.

\bibliographystyle{ws-procs961x669}
\bibliography{biblio}

\end{document}